\documentclass{article}

\usepackage{pifont}
\usepackage{amssymb}
\usepackage{microtype}
\usepackage{graphicx}
\usepackage{subfigure}
\usepackage{booktabs}

\usepackage{comment}

\usepackage{hyperref}
\usepackage{xurl}

\usepackage[accepted]{icml2024}

\usepackage{amsmath}
\usepackage{amssymb}
\usepackage{mathtools}
\usepackage{amsthm}
\usepackage{flushend}
\usepackage{float}

\usepackage[capitalize,noabbrev]{cleveref}

\theoremstyle{plain}

\theoremstyle{definition}

\theoremstyle{remark}

\usepackage[textsize=tiny]{todonotes}

\icmltitlerunning{The Case for ESM3 as a General-Purpose AI Model with Systemic Risk Under the EU AI Act}

\begin{document}

\twocolumn[
\icmltitle{The Case for ESM3 as a General-Purpose AI Model with Systemic Risk Under the EU AI Act}

\icmlsetsymbol{equal}{*}

\begin{icmlauthorlist}

\icmlauthor{Taro Qureshi}{equal,sch_taro}
\icmlauthor{Jacob Griffith}{equal,sch_jacob}
\icmlauthor{Koen Holtman}{comp_aisl}
\icmlauthor{Marcel Mir Teijeiro}{comp_aisl}
\icmlauthor{Ze Shen Chin}{comp_aisl,comp_om}
\icmlauthor{Rokas Gipiškis}{comp_aisl,rokas_uni}

\end{icmlauthorlist}

\icmlaffiliation{comp_aisl}{AI Standards Lab}
\icmlaffiliation{comp_om}{Oxford Martin AI Governance Initiative}
\icmlaffiliation{rokas_uni}{Vilnius University}
\icmlaffiliation{sch_taro}{Northeastern University London}
\icmlaffiliation{sch_jacob}{London School of Economics and Political Science}

\icmlcorrespondingauthor{Taro Qureshi}{tarojq@gmail.com}
\icmlcorrespondingauthor{Jacob Griffith}{jacobglyngriffith@outlook.com}
\icmlcorrespondingauthor{Ze Shen Chin}{zeshen@aistandardslab.org}
\icmlcorrespondingauthor{Rokas Gipiškis}{rokas@aistandardslab.org}

\icmlkeywords{Biorisk, Biological Tools, AI Governance, AI Safety, European Union}

\vskip 0.3in
]

\printAffiliationsAndNotice{\icmlEqualContribution} 

\begin{abstract}
Due to ambiguity in the wording of the EU AI Act, we examine the question of to what extent frontier biological foundation models such as ESM3 are subject to obligations for general-purpose AI models with systemic risk under the EU AI Act. In this paper, we map ESM3 to the biorisk chain, and conclude that it would be desirable if the providers of ESM3 and similar biological models were subject to these obligations, which would require them to assess and mitigate dual-use risks from their models. We then perform an analysis, comparing the attributes of ESM3 to the classification criteria in the AI Act and the supporting material. We conclude that at this time, ESM3 does not appear to be meaningfully regulated by the Act. We then propose remedies to correct the situation.

\end{abstract}

\section{Introduction}
\label{Introduction}
AI and biotechnology are converging into software tools designed to streamline or automate several fundamental processes in biological science, such as protein design and structure prediction. These are referred to as biological tools (BTs). While having a great number of positive applications, such as vaccine and treatment design, concerns have been raised by researchers that BTs may also have the potential to uplift the capabilities of bad actors who wish to cause harm using biological weapons, for example by making the process of designing novel pathogens easier. 

The EU AI Act \cite{noauthor_eu_nodate} is the world’s first comprehensive legal framework aimed at addressing the risks associated with AI within the European Union. Part of this framework classifies models that meet specific criteria as general-purpose AI (GPAI), which may or may not also be a general-purpose model with systemic risk (GPAISR). Receiving a designation of GPAI is a prerequisite for a model being classified as presenting systemic risks. This then creates a regulatory blind spot wherein a model may present systemic risk, but not meet the underlying narrow criteria for being classified as a GPAI model, and thereby remain exempt from meaningful safety regulation, despite its potential to be used to cause harm.  

We focus on the frontier biological tool Evolutionary Scale Modeling version 3 (ESM3) and explore how it exists within this regulatory blind spot. ESM3, a family of models of various sizes, is chosen because, at the time of writing, it is one of the most prominent biological foundation models, with an open-source and thus easily accessible variant. 

The EU's approach to regulation is critical as it is likely to influence other major jurisdictions in developing their own governance frameworks, as per the ``Brussels Effect" \cite{bradford_2012_brussels}.

Our contributions are: (1) a qualitative mapping of ESM3's capabilities to the biorisk chain; (2) demonstrating that the regulatory blind spot has two reinforcing sources: in the Act's ambiguous definitions and in the Commission's General-Purpose AI Guidelines (henceforth GPAI Guidelines); and (3) a discussion of four potential policy remedies.

\section{Related Work}
 
An AI-powered biological disaster is one of the most commonly proposed pathways to catastrophic risk scenarios \cite{nelson_cassidy_understanding_2023}.
The real-world impact of machine learning-enabled biological tools on biorisk is a nascent field of study, and researchers disagree about their potential to contribute to harm. 

Peppin et al. \citeyearpar{peppin_reality_2025} have proposed that while contemporary biological tools are highly capable, they are unlikely to contribute to the uplift of real-world biorisk. They have also stated that risk assessments are difficult, as there are relatively few robust biorisk models available. 

Others argue that it is highly likely that AI tools will significantly uplift the capability of bad actors to develop chemical or biological weapons \cite{brent_contemporary_2025, luckey_mitigating_2025}.

\citet{nelson_cassidy_understanding_2023} highlighted the potential for regulatory loopholes to be deliberately exploited by developers as early as 2023, citing the dangers of narrowing the scope of models to which oversight and regulatory measures apply. They also developed a biorisk chain model, which is useful for explaining the impact of BTs on biorisk uplift in terms of specific steps in a longer process, enabling more detailed risk evaluation. Their work was published before the \textit{Guidelines on the Scope of the Obligations for General-Purpose AI Models} \cite{EC_GPAI_2025} were published. We develop their analysis further in light of the publication of these GPAI Guidelines.

Moulange et al. \citeyearpar{moulange_capability-based_2025} have conducted analyses on specific capabilities of frontier BTs, connecting them to specific steps in the biorisk chain. Most directly, \citet{moulange_biological_2025} examined BTs under the EU AI Act and concluded that the current framework struggles to accommodate domain-specialist foundation models. Our contribution extends their analysis through a single-model case study grounded in ESM3's documented capabilities and a detailed reading of the Commission's GPAI Guidelines published after their work. We make use of both of these contributions to add empirical data to our analysis of the EU AI Act.

\citet{hopkins_biological_2025} has argued that the European Union's new AI regulations fail to regulate highly capable and potentially dangerous biological models. Our work aims to investigate this claim by examining the specific family of frontier biological tools, ESM3, and directly mapping capabilities to the biorisk chain. 

While we focus on ESM3, the regulatory gap we identify can potentially apply more broadly to biological foundation models, including AlphaFold \cite{jumper2021highly}, RFdiffusion \cite{watson2023novo}, and ProtGPT2 \cite{ferruz2022protgpt2}, none of which interact in human language, despite having broad use-cases.

\section{ESM3 - Overview}

ESM3 is the first of its kind able to simultaneously reason over three fundamental properties of proteins:

\begin{itemize}
    \item \textbf{Sequence}: The linear chain of amino acids that makes up a protein.
    \item \textbf{Structure}: The three-dimensional shape of the protein that determines its function.
    \item \textbf{Function}: The biological role or activity of the protein.
\end{itemize}

An ESM3 model analyses and learns patterns using a process of tokenisation similar to that of a Large Language Model (LLM), and can ``fill in” missing sequence, structure, or function data. A user prompts the model by providing a mix of known and unknown information. The unknown parts are ``masked," and the model predicts them. This could be a partial amino acid sequence, which the model would complete, or keywords relating to the function of the desired protein. The ESM3 model then returns a protein, represented as a sequence of amino acids, or a set of 3D atomic coordinates that describe its predicted structure. In this way, a user can iteratively design novel proteins \cite{hayes_simulating_2024}.

This represents a significant leap forward in what is possible, as a user can now test thousands of variants computationally in a fraction of the time it would take in a traditional lab, also allowing for potentially unexpected insights and controlling for stability, countermeasure evasion (such as vaccines or other infection control protocols) and other characteristics.

The ESM3 model is available in three sizes: small (1.4B parameters), medium (7B parameters) and large (98B parameters). Its 1.4B-parameter open variant is freely available on HuggingFace and GitHub, whereas the medium and large variants are not open-source and can only be accessed via API. Within this work, we rely on ESM3-Open (simply referred to as ESM3) as it is the easiest model for a bad actor to clone and run on their machine \cite{halstead_managing_2024}.

The code and model weights for ESM3 are available under a mixture of non-commercial and permissive commercial licenses, and there are multiple scales and variants of models available, with the smallest being trained with 1.07$\times{10^{24}}$ FLOPs of compute on 771 billion tokens, with 1.4 billion parameters \citep{hayes_simulating_2024}. 

The 1.4B parameter version of ESM3 is free for academic and research purposes and is hosted on HuggingFace and Github. This makes the software accessible even to those with low technical proficiency. However, the larger 7B and 98B parameter variants are gated behind EvolutionaryScale's Forge API, which includes safety filtering. Whilst the larger models are expected to perform better on harder tasks, the core functionality is the same across all three variants.

\section{Biorisk Chain}
\label{biorisk_chain}

\begin{figure*}[ht]
    \centering
    \includegraphics[width=\textwidth]{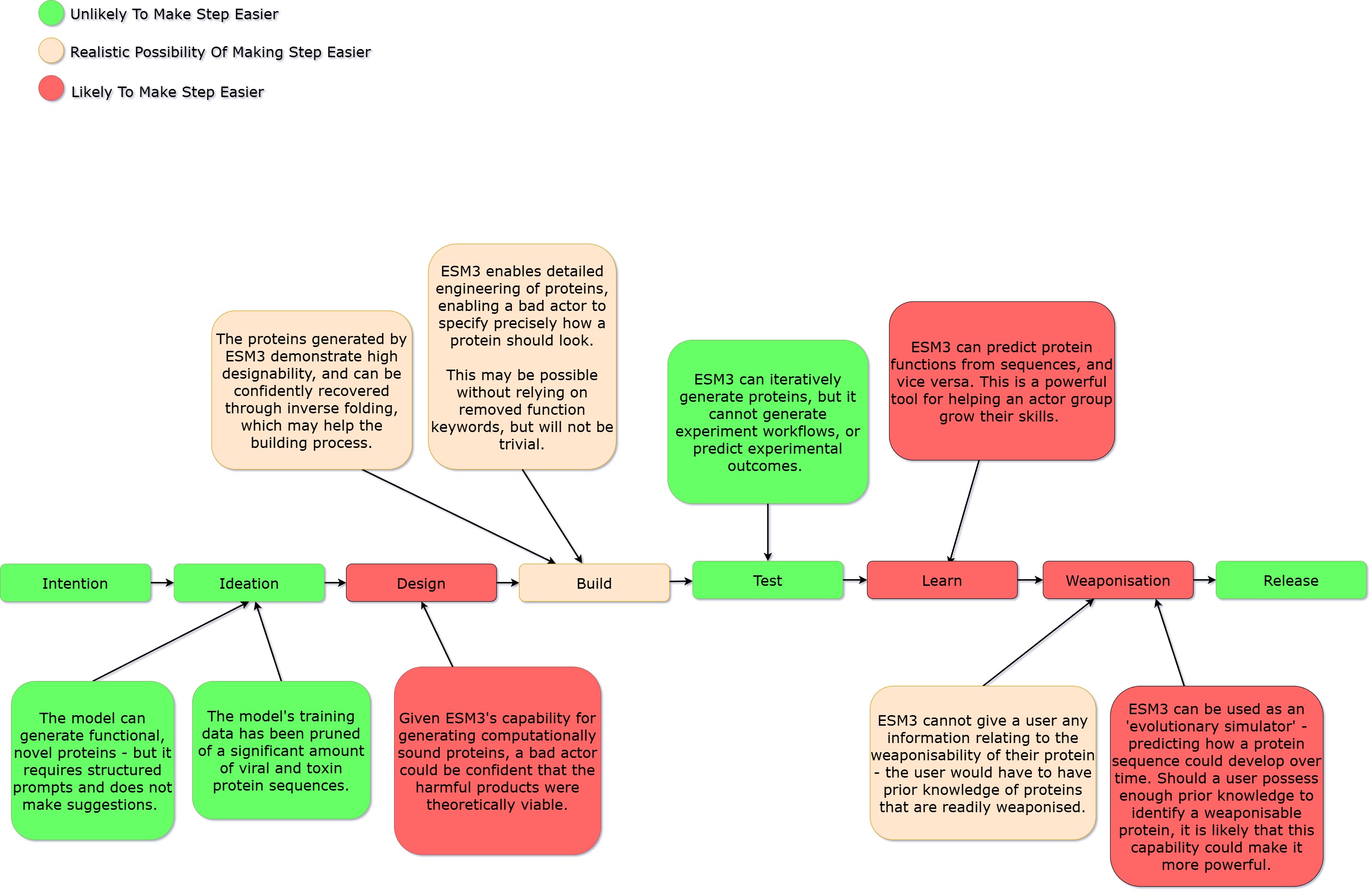}
     \caption{ESM3’s Capabilities and the Biorisk Chain}
        \label{fig:fig_1}
\end{figure*}

The biorisk chain is a tool for illustrating the series of steps a malicious actor must successfully complete to carry out a biological attack. The main impact of the availability of new technologies is to make steps easier, or to enable some steps to be bypassed altogether. Sandberg \& Nelson \citeyearpar{sandberg_who_2020} conducted a prior study mapping commonly displayed BT capabilities to steps in the chain. Using this as a basis, we select for ESM3's specific capabilities and map them to the biorisk chain, to illustrate the specific steps that will be made easier by BTs such as ESM3. 
\\

The chain consists of seven steps corresponding to the series of tasks or processes a malicious actor or group of actors must successfully complete before committing an act of harm, beginning with having the intention to undertake a biological attack, developing the idea for the specific pathogen or agent that could be used, designing and testing the agent in silico or in a real world lab, weaponising the agent and ending with releasing it into the target population. We then make a subjective judgement on the plausibility of each step in the biorisk chain being made easier by an actor with access to an ESM3 model, given the functions and capabilities of the model. \footnote{To avoid giving a sense of objectivity to judgements about which there is a significant level of uncertainty, we relied on the framework employed by the Professional Head of Intelligence Assessment, the Probability Yardstick \citep{noauthor_explaining_nodate}, organising our analyses into the categories of ``likely", ``realistic possibility" and ``unlikely".} 

While \citet{peppin_reality_2025} have held that for BTs to meaningfully contribute to real-world biorisk uplift, they must unilaterally affect the chain, \citet{sandberg_who_2020} have demonstrated mathematically that, assuming biorisk events and actors follow a power-law distribution, making a number of steps in the biorisk chain easier increases the possibility of success, as the length of the chain is effectively reduced. As per the products of our analyses presented in \autoref{fig:fig_1}, of the seven steps in the standard chain, four of those steps (Design, Learn, Build and Weaponisation) can potentially be made easier with access to an ESM3 model. 
\\
This is particularly concerning given the open-source accessibility of ESM3-Open, the open-source variant of ESM3, meaning actor groups need not possess considerable resources to make use of the tool. This analysis lends weight to the critical importance of regulation at the software level. The population of lower-power actor groups is diffuse and highlights the necessity for regulation that is pro-active and dynamic.

\section{Regulatory Scope of the EU AI Act} %

The Act has been designed in order to mitigate the worst AI-associated harms while also allowing for innovation in developing systems with many beneficial use-cases. Article 55 of the Act obligates providers of ``general-purpose AI models with systemic risk" to assess and mitigate possible risks that may come from the release of their model. This includes mitigating the dual-use risk where the release of the model will lower barriers to entry for the development of bioweapons. A typical mitigation would be to restrict who can access the model. For any access-restriction obligations to apply to ESM3, it would have to be classified as a general-purpose AI model with systemic risk. We show below that this does not seem to be possible within the current regulation. ESM3 seems to fall into a blind spot created by two reinforcing sources: ambiguities in the Act's statutory text, and a narrowing interpretation in the Commission's non-binding GPAI Guidelines.

\subsection{The Act's Definition and Its Ambiguities}
Chapter 3 of the EU AI Act provides rulings for the classification of AI systems as high-risk. For a model to be considered to possess systemic risk, it must first meet the criteria to qualify as ``general-purpose AI" model. The Safety and Security chapter of the Code of Practice for General-Purpose AI Models (Appendix 1.3.1 and Appendix 1.4) \cite{CoPGPaiSafetyChapter} places CBRN risks at the core of what the legislators consider systemic risk, and therefore must be assessed and mitigated by signatories. As shown in Section \ref{biorisk_chain}, there is a possibility that an ESM3 model contributes to potential uplift in the design and weaponisation stages of biological weapons development — particularly in testing sequences and structures with dangerous functions and predicting how a weaponised pathogen might evolve to evade environmental detection (see Section \ref{biorisk_chain}). This wording, particularly from the Code of Practice signals clear intent to mitigate risks of the given sort - however, the wording of the subsequent Articles and supporting Recitals work against this. 

Article 3(63) of the EU AI Act \cite{noauthor_eu_nodate} defines a general-purpose AI model as one that displays ``significant generality" and ``is capable of competently performing a wide range of distinct tasks", and that ``can be integrated into a variety of downstream systems or applications". 

As the third criteria is plainly met, we can start here. ESM3 can certainly be integrated into a variety of downstream systems: it can produce biologically meaningful representations and predictions that other software can use for ranking, design, monitoring, or decision support. ESM3 is described as supporting mutation screening and stability/function optimization, as well as it can generate candidate protein variants, then pass them into a downstream screening and wet-lab validation pipeline, as in the esmGFP protein-design case \cite{hayes_simulating_2024}.

When it comes to ``generality", we turn to the recitals to glean the legislative intent. Recital 97, one of the few places that the word ``generality" appears in the Act, sheds no light onto its meaning within the Act. It simply restates that GPAI classification depends on generality and the ``capability to competently perform a wide range of distinct tasks."

Recital 98 states that ``whereas the generality of a model could, inter alia, be determined by a number of parameters, models with at least a billion parameters and trained with a large amount of data using self-supervision at scale should be considered to display significant generality and to competently perform a wide range of distinctive tasks." ESM3, with 1.4 billion parameters, exceeds this indicator. It could therefore be argued that Recital 98 opens the door to consider ESM3 a model with ``generality" under the AI Act.

Yet Recital 99 gives multi-modal models involving text, audio, images or video as key examples of models that could ``accommodate a wide range of distinctive tasks", suggesting that the legislators had a very broad range of capabilities, coming from the inclusion of human language in the model, in mind.

The Act's text related to general-purpose AI models is thus ambiguous with respect to biological models. ``Significant generality" is undefined: the Act does not specify whether domain-internal generality is sufficient. It is unclear whether ESM3's tasks (de novo sequence generation, structure prediction, inverse folding, functional annotation, and iterative protein design) could qualify as a ``wide range of distinct tasks". 

Article 3(65) of the Act defines ``systemic risk”  as a risk specific to the high-impact capabilities of GPAI models, with ``reasonably foreseeable negative effects on public health, safety, public security, fundamental rights, or the society as a whole.” Recital 110 of the Act elaborates on such high-impact capabilities, citing chemical, biological, radiological, and nuclear (CBRN) risks and noting that such risks include ``the ways in which barriers to entry can be lowered, including for weapons development.” So unlike recital 99, this recital does suggest the legislator was thinking beyond models that involve human language.

Tying the systemic risk provision to a designation of general-purpose that is itself contingent on human language modeling seems to hamstring this mechanism and obstruct its intent. This is made clear by Recital 110’s explicit reference to ``lowering barriers to entry […] for weapons development''. As demonstrated in Figure~\ref{fig:fig_1} above, an ESM3 model could provide malicious actors with the following:

\begin{itemize}
\item More efficient testing processes

\item Insight into why certain development stages succeed or fail

\item Reliable production of relevant biological components at larger scales
 \end{itemize}
 
Those with sufficient biological expertise could deploy the model for such purposes (and jailbroken LLMs could further bridge the knowledge gap, equipping novices with much of what is needed save access to a physical lab and materials) making the model a perfect example of a tool that could be used to lower barriers to entry for weapons development by making one or more elements of the process easier.

Therefore, ESM3 meets or exceeds many of the criteria that the Act associates with systemic risk. The doubt relies on the interpretation of the terms ``significant generality"  and ``human language" to be classified as a GPAI model.

\subsection{The GPAI Guidelines' Interpretation}

The Commission's GPAI Guidelines \cite{EC_GPAI_2025} offer non-binding interpretive guidance of the AI Act.  
Paragraph 9 of the GPAI Guidelines states explicitly that an authoritative interpretation of the Act ``may only be given by the Court of Justice of the European Union". Paragraph 9 concludes that ``a case-by-case assessment will always be necessary to account for the specifics of each individual case." Despite this flexibility, the GPAI Guidelines resolve the GPAI classification ambiguity in a direction that excludes biological models.

\subsubsection{GPAI Consideration}

Paragraph 17 of the GPAI Guidelines states that the ``indicative criterion for a model to be considered a general-purpose AI model is that its training compute is greater than \(10^{23}\) and it can \textbf{generate language} (whether in the form of text or audio, text-to-image or text-to-video).'' 

Even the smallest variant of ESM3 exceeds the compute threshold by an order of magnitude, at $1.07 \times 10^{24}$ FLOPs. However, ESM3 does not interact with users in the form of human language, and if this is how ``generate language" above is understood, then ESM3 cannot be considered general-purpose according to Paragraph 17. This would then block this pathway for ESM3 and many other similarly powerful BTs to be considered GPAI models.

In Paragraph 19, the GPAI Guidelines state that these specific modalities have been ``chosen based on the fact that models trained to generate language, whether through text or speech (as a type of audio) are able to use language to communicate, store knowledge and reason. No other modality confers such a wide range of capabilities.'' The specific reference to ``speech (as a type of audio)'' suggests that language in the context of the GPAI Guidelines refers to human language, and the modalities thought to be indicators of generality have been thus chosen. The qualities Paragraph 19 identifies, such as the capacity to communicate with human beings, store knowledge in human-accessible form, and reason in a manner legible to humans, are intrinsically tied to human language as a modality.

Whilst not human language, ESM3 nevertheless operates across three \textit{protein} modalities: sequence, structure, and function. Through tokenisation, it can process these modalities in the same latent space, storing large-scale biological knowledge and reasoning over patterns in this data to simulate evolutionary pathways. ESM3's generality lies not in crossing domains beyond biology, but in its capacity to perform a wide range of protein design and prediction tasks with minimal task-specific retraining. These modalities also signal powerful and diverse downstream integration capability as protein strings are a fundamental abstraction used in most biological use-cases. 

The GPAI Guidelines further provide examples of types of models considered out of scope, which closely match the profile of ESM3: ``A model trained for playing chess or videogames [...]. A model trained specifically for modelling weather patterns or \textbf{physical systems}." \footnote{Emphasis ours.} 

Text-to-image and text-to-video modalities are mentioned in Paragraph 19 of the GPAI Guidelines as narrow use-cases that may permit a model to be considered general-purpose, but notably, the input must still be in the form of human language. ESM3 receives no human-language input at any stage.

The GPAI Guidelines even narrow scope within human language, stating that ``a model that can generate speech in the form of song lyrics [will not be considered general-purpose] if it can only competently perform a narrow set of tasks". If models with narrow human-language capabilities do not qualify, a model with no human-language capabilities plainly cannot.

\subsubsection{General-Purpose AI with Systemic Risk (GPAISR) Consideration}

In Section 2.3.1 of the GPAI Guidelines, the Commission recalls that the legislator has set a compute training threshold of \(10^{25}\) FLOPs, above which a GPAI is automatically classified as a GPAISR. ESM3, trained on \(10^{24}\) FLOPs, falls below this automatic threshold. However, Article 51 states that:
``A general-purpose AI model shall be classified as a general-purpose model with systemic risk if it meets \textit{any} [emphasis ours] of the following conditions […]”. 

\begin{itemize}
\item Article 51(1)(a): It has high impact capabilities evaluated on the basis of appropriate technical tools and methodologies, including indicators and benchmarks.
\item Article 51(1)(b): Based on a decision of the Commission, ex officio or following a qualified alert from the scientific panel, it has capabilities or an impact equivalent to those set out in point (a) having regard to the criteria set out in Annex XIII.
\item Article 51(2): The cumulative amount of computation used for its training measured in floating point operations is greater than $10^{25}$.

\end{itemize}

While ESM3 falls short of the compute threshold, it has other qualifying properties:

\begin{itemize}

\item Specific input and output type.  What is very notable is that biological sequences are explicitly listed in Annex XIII as an example type. This indicates that the legislator had biological risks in mind when developing Annex XIII.

\item State-of-the-art capability – Peer-reviewed work presents ESM3 as a frontier multimodal generative model for protein design \cite{hayes_simulating_2024}.
\end{itemize}

Thus, Article 51(1)(b) of the Act enables a decision from the Commission designating ESM3 as GPAISR, despite it being below the outlined compute threshold. Such a decision needs to be motivated by the observation that the model has high-impact capabilities. We have discussed these high-impact capabilities in the form of uplift on certain stages of the biorisk chain in Section \ref{biorisk_chain}. 

\section{Potential Remedies to Clarify the Status of ESM3 and Similar Models}
In this section, we briefly review some non-mutually-exclusive remedies to the identified regulatory ambiguity that currently exists.

\subsection{Use Article 3(63) and Article 51(1)(b) to designate ESM3 as a GPAI model with systemic risk}

The Commission could directly rely on Article 3(63) to designate ESM3 as general-purpose AI model, by using a broader interpretation of terminologies such as ``generality'' and ``distinct tasks''. Once it is considered GPAI, the Commission can then use Article 51(1)(b) to designate ESM3 as a GPAI model with systemic risk due to its ``high impact capabilities''. It would have to publish this decision somewhere, while also informing the providers of ESM3. Preferably, the Commission should also revise the GPAI Guidelines document soon after.

\subsection{Revise the GPAI Guidelines}
The Commission can revise Section 2.1 of the GPAI Guidelines to broaden the modality criterion. This could take the form of removing the language requirement entirely, adding a clear exemption to this requirement or introducing a domain-generality test that captures models performing a wide range of distinct tasks within a specialised domain (such as protein biology). Since the GPAI Guidelines are non-binding interpretive guidance, such a change can be done by the Commission without the involvement of other actors. 

\subsection{Amend the Act}

An amendment could decouple systemic-risk designation from GPAI status, by amending Article 51 to allow models meeting Annex XIII criteria to be designated directly, regardless of whether they qualify as GPAI. This is robust, but requires a lengthy legislative process.

\subsection{Complementary EU Instruments}
The AI Act might not be the only available instrument relevant to software-enabled biorisk uplift. The EU's dual-use export control regime (Regulation (EU) 2021/821) covers certain biotechnology items but does not currently list biological AI model weights or software. Adding biological foundation models to the dual-use control list would address cross-border proliferation risks that the AI Act's market-focused framework is not designed to catch \cite{cfg_priorities_2025}, but this must still be developed further to mitigate risk arising from bad actors within the Union itself.

Additionally, the EU's broader biosecurity policy initiatives, including the Council's 2025 priorities on European biosecurity, could incorporate AI-specific provisions requiring risk assessment for BTs, creating domain-specific regulation that complements the AI Act's horizontal framework, such as the proposed EU Biotech Act \cite{EU2025BiotechActProposal}. These initiatives are in the early stages and the actual effects remain to be seen.

\section{Limitations}

The biorisk chain model, while useful, suffers from the fact that it is a simplification of an incredibly complex world. The practical barriers of acquiring materials, laboratory access, or avoiding detection, which remain significant obstacles even with computational tools, are abstracted into a single step. Furthermore, our risk ratings for each step of the biorisk chain are based on subjective and qualitative assessment using ESM3's documentation, and they are not validated through empirical measurement. 

\section{Conclusion}

This paper has examined the regulatory gap in which ESM3 and other similar BTs currently exist.

ESM3 does not seem to meet the criteria for GPAI classification under the current wording of the GPAI Guidelines. These leave little room for interpretation of what ``language" refers to outside of human language, and given that this is key to what is considered general-purpose in the GPAI Guidelines, they offer no clear regulatory pathway for ESM3 and other BTs to be classified as GPAI. This would then mean it will not automatically be subject to the obligations for meaningful risk-mitigation in Article 55 of the Act.

The Act, as expressed through recitals, aims to leave room to regulate systems that pose CBRN risks. However, the over-reliance on natural language as a signifier of generality in the GPAI Guidelines hampers the effectiveness of the systemic risk provision, limiting it effectively to LLMs and some forms of generative AI. 

We propose several potential remedies to the current situation of unwanted ambiguity. The use of Article 51(1)(b) is the easiest short-term remedy, but we prefer that this remedy is combined with an updating of the GPAI Guidelines.

\section*{Impact Statement}

The goal of this work is to advance the field of AI safety and improve the governance of frontier biological tools. There are many potential societal consequences of our work, including improved regulatory clarity and stronger oversight of capable biological models, as well as the possibility that our analysis could be used to inform future policy debates or system design choices in ways we do not intend. We believe the benefits of identifying and reducing regulatory gaps for high-capability biological models outweigh these risks, but we encourage careful, responsible use of the ideas presented here.

\section*{Acknowledgements}

We thank Alicia Pollard for practical support.

\bibliographystyle{icml2024} 
\bibliography{aaai2026}      
\end{document}